%% file: paper.tex
\documentclass[sigconf]{acmart}

\usepackage{booktabs} 
\usepackage{graphicx}
\usepackage{subcaption}

\setcopyright{rightsretained}

\acmConference[KDD'18 Deep Learning Day]{ACM KDD'18 Deep Learning Day}{August 2018}{London, UK}
\acmYear{2018}
\copyrightyear{2018}

\begin{document}
\title{Five lessons from building a deep neural network recommender for marketplaces}

\author{Simen Eide}
\affiliation{%
  \institution{Schibsted Media Group}
  \city{Oslo} 
  \country{Norway}}
\email{simen.eide@schibsted.com}

\author{Audun M. {\O}ygard}
\affiliation{%
  \institution{Schibsted Media Group}
  \city{Oslo} 
  \country{Norway}}
\email{audun.oygard@schibsted.com}

\author{Ning Zhou}
\affiliation{%
  \institution{Schibsted Media Group}
  \city{Oslo} 
  \country{Norway}}
\email{ning.zhou@schibsted.com}

\renewcommand{\shortauthors}{S. Eide, A. M. {\O}ygard, and N. Zhou}

\begin{abstract}
Recommendation algorithms are widely adopted in marketplaces to help users find the items they are looking for. The sparsity of the items by user matrix and the cold-start issue in marketplaces pose challenges for the off-the-shelf matrix factorization based recommender systems. To understand user intent and tailor recommendations to their needs, we use deep learning to explore various heterogeneous data available in marketplaces. This paper summarizes five lessons we learned from experimenting with state-of-the-art deep learning recommenders at the leading Norwegian marketplace \textit{FINN.no}. We design a hybrid recommender system that takes the user-generated content of a marketplace (including text, images and meta attributes) and combines it with user behavior data such as page views and messages to provide recommendations for marketplace items. Among various tactics we experimented with, the following five show the best impact: staged training instead of end-to-end training, leveraging rich user behaviors beyond page views, using user behaviors as noisy labels to train embeddings, using transfer learning to solve the unbalanced data problem, and using attention mechanisms in the hybrid model. This system is currently running with around 20\% click-through-rate in production at \textit{FINN.no} and serves over one million visitors everyday.
%
%
%
%
\end{abstract}

%
%

\ccsdesc[500]{Information systems~Content ranking}
\ccsdesc[300]{Applied computing~E-commerce infrastructure}

\keywords{recommender system, deep learning, marketplace}

\maketitle

\input{body-conf}

\bibliographystyle{ACM-Reference-Format}
\bibliography{bibliography} 

\end{document}

%% file: body-conf.tex
\section{Introduction}
\label{sec:intro}

\begin{figure}
\centering
\includegraphics[width=0.45\linewidth]{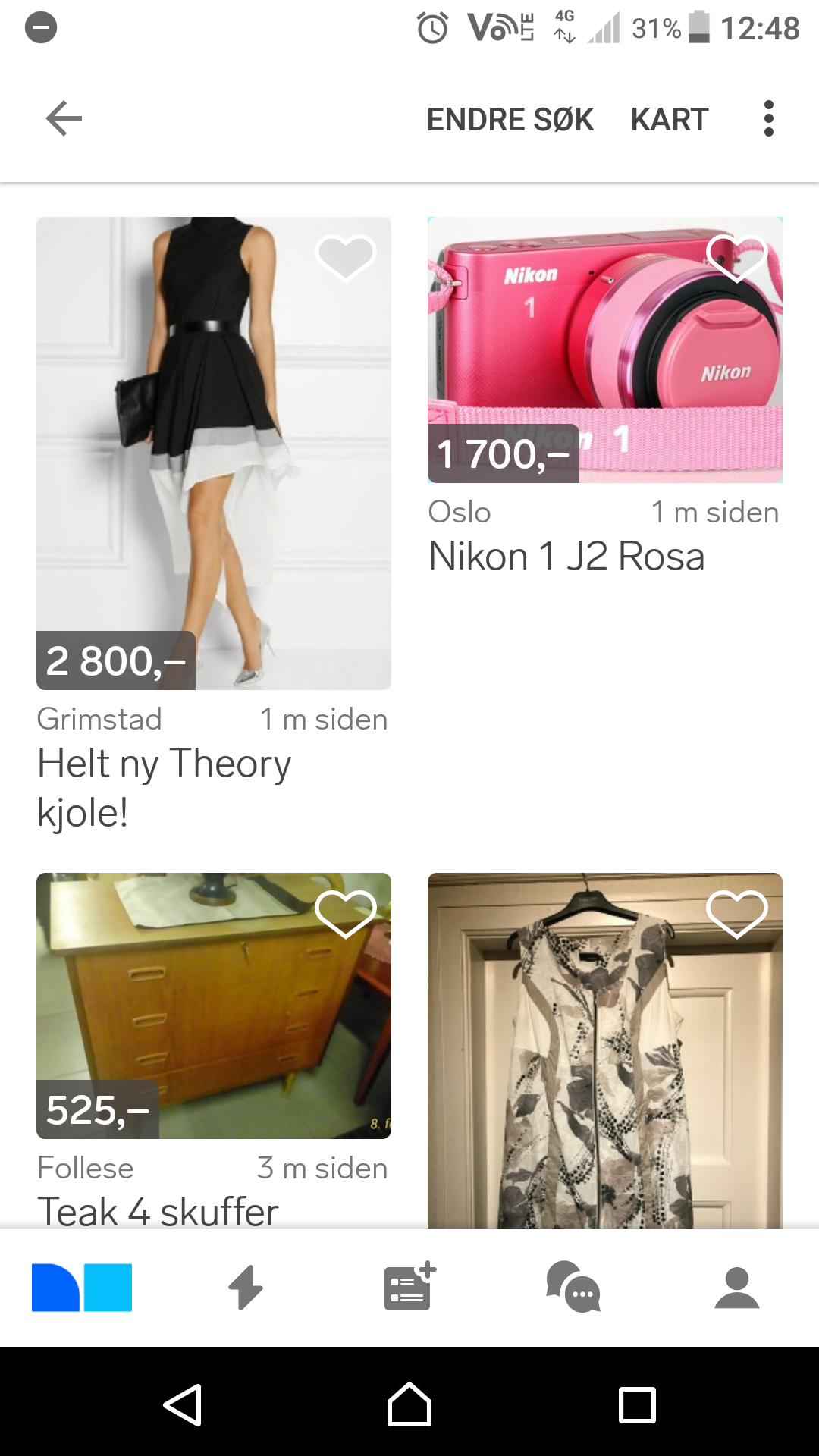}
\hspace{0.02\linewidth} 
\includegraphics[width=0.45\linewidth]{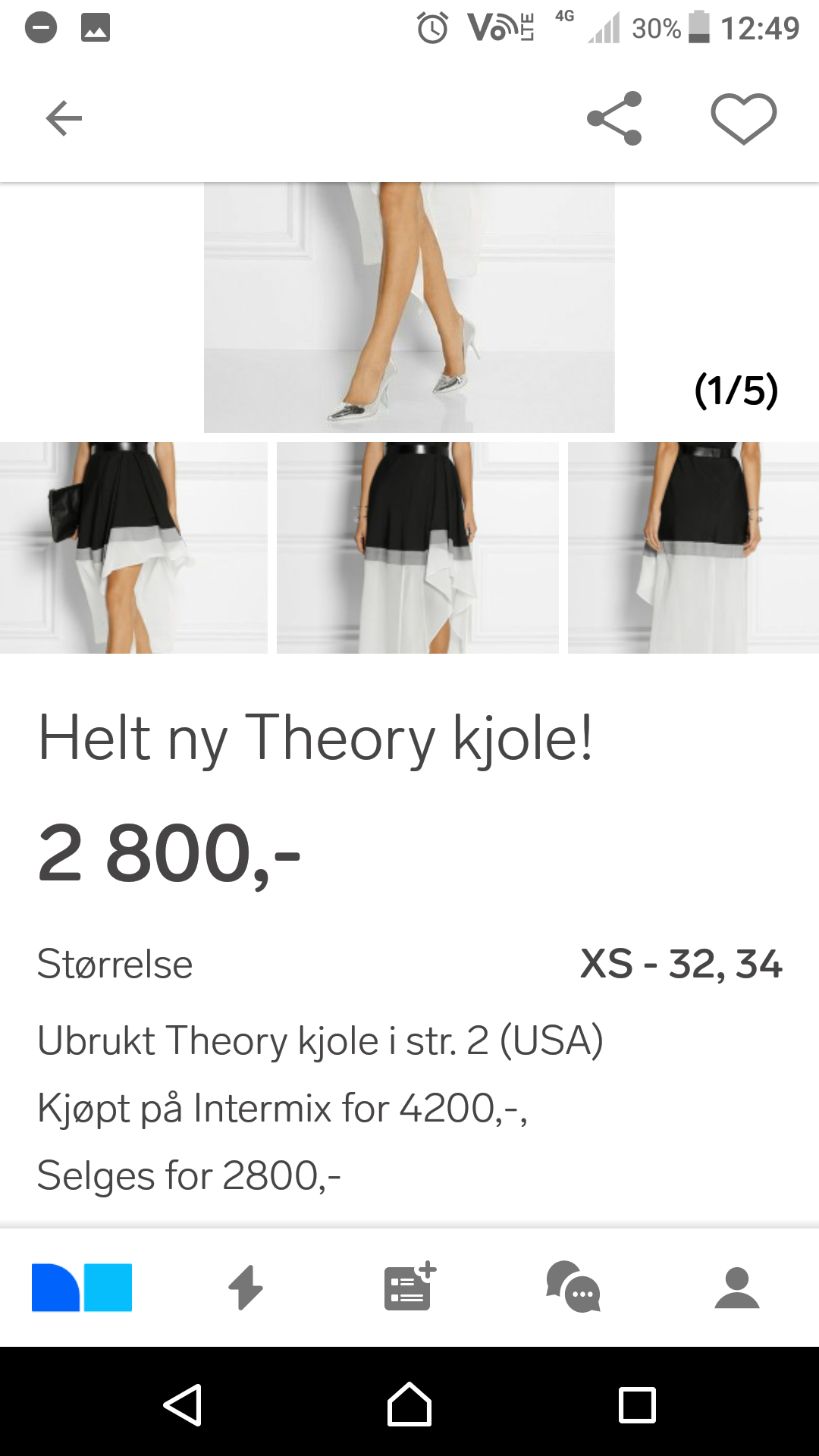}
\caption{An example of \textit{FINN.no} marketplace. It shows the screen shot of a search result page on the left and an ad detail page on the right.}
\label{fig:marketplace}
\end{figure}

The common business model of marketplaces is to provide an online platform where sellers post their items for sale and buyers search for items they want to purchase. The items can range from low-value ones such as secondhand textbooks and clothes to high-value ones such as cars and real estate properties. Sellers can also post non-tangible items such as job opportunities. In the marketplace industry, a \textit{classified} is a post from a seller selling one or multiple items, and we refer to it as an "ad" in this paper. An example is shown in Figure \ref{fig:marketplace} with a search result page on the left and an ad detail page on the right. Marketplaces can have multiple verticals, but there are often specialized marketplaces on house, car and job to adapt to the specific characteristics of those categories. In this paper, we use \textit{FINN.no} (\url{https://finn.no/}), the leading online marketplace in Norway, to understand the user-content interaction mechanism on marketplace and to experiment our recommender system with real users. It is a multi-vertical marketplace, but for the work described here, we focus only on the \textit{torget} vertical that contains mostly secondhand daily items such as books, clothes, furnitures and pets.

In essence, the \textit{FINN.no torget} marketplace can be viewed as a special type of e-commerce site that provides secondhand items across many different categories from a very large and scattered non-professional seller community. Recommendation for marketplace ads is more challenging than the standard e-commerce product recommendation for the following reasons: 1) Many marketplace sellers are non-professional individuals, and the information they provide to describe the items for sale is often of lower quality (incomplete or inaccurate) and less standardized. Thus it is more challenging to solely rely on the content-based features to identify similar items. 2) Location proximity is often neglected in the off-the-shelf recommender solutions, but it actually plays an important role in marketplaces, since the potential buyers often prefer to complete the transaction in person to check if the item fits their expectation. 3) Different from e-commerce sites where each product usually has abundant supply, marketplace items are often secondhand and therefore unique in some sense. As a result, item volatility in marketplaces is often much higher than in the e-commerce scenario. While in traditional recommendation scenarios the number of items is much smaller than the number of users, in marketplaces the number of "active" items has the same order of magnitude of the number of users. 4) \textit{FINN.no} does not use the auction method, and most sellers reserve their items for the first buyer whom they reach a deal with. Therefore freshness is also a factor to take into account in recommendation solutions, so that potential buyers can find newly listed items sooner. This highlights the importance of solving the cold start problem for newly listed items.
%
%
%
%
%

We experimented with a group of hybrid recommenders in production at \textit{FINN.no} to solve the challenges mentioned above. Among various things we tried, the following five have shown significant improvements in the A/B tests with users: rich user behavior signals in Section \ref{sec:behavior}, user behavior trained embeddings in Section \ref{sec:embedding}, transfer learning for images in Section \ref{sec:image}, staged training strategy in Section \ref{sec:staged}, and attention mechanism in Section \ref{sec:attention}. 

\section{Related work}
\label{sec:related}
%
%
Search and discovery are the two main ways for users to find relevant items on a content platform. Using recommender systems to improve the discovery experience has been a hot topic in recent years. Both collaborative filtering \cite{barkan2016item2vec} \cite{sarwar2001cf} and content based methods \cite{grbovic2015commerce} are commonly used in product ranking for e-commerce. Recent works like LightFM \cite{Kula15:lightfm} combine the two to address the cold-start problem for personalized recommendations. The state-of-the-art recommenders such as \cite{Criteo:2016} and \cite{pinterest:2017} often use learning-to-rank to model from a complex set of features including text, image, category, user profile, user behavior, etc. Models of cascade \cite{liu2017cascade} or sequential attention \cite{atrank:2018} that consider a longer user behavior history also show good results. The use of multi-armed bandits to prioritize from multiple sources is discussed in different industries. For example, Pinterest \cite{StephaniedeWet17} uses a blending type of algorithm to generate a feed mixing their recommendation content and social content. We draw inspiration from previous work and experiment with hybrid recommenders that leverage the behavior-content complementation.

\section{Recommender overview}
\label{sec:exp}

\begin{figure*}
\centering
\includegraphics[width=\linewidth]{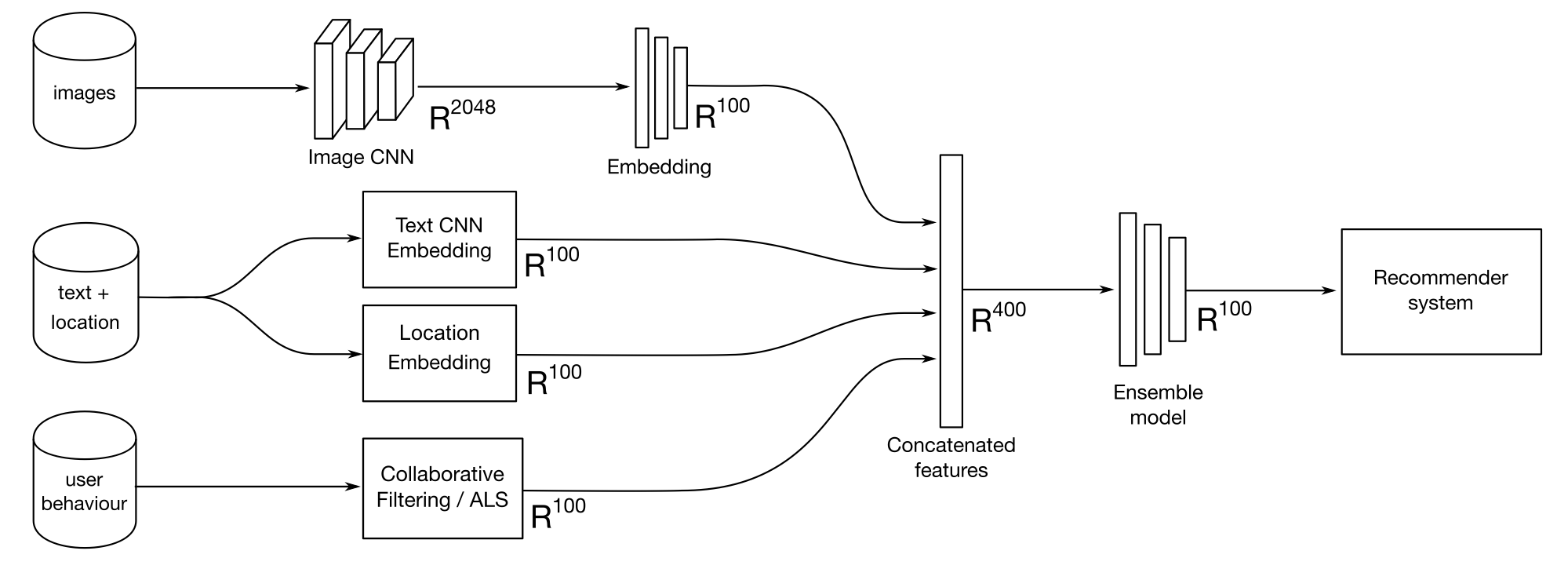}
\caption{Recommender system overview.}
\label{fig:overview}
\end{figure*}

We start with an off-the-shelf matrix factorization model as the baseline. In addition to the classical cold start problem, the particular challenges for marketplaces mentioned in Section \ref{sec:intro} make it even harder for the off-the-shelf solution to perform. Similar to Lightfm\cite{Kula15:lightfm}, we look into combining content-based features with user behavior signals to solve the cold start problem. We experimented with a group of hybrid models using item representations from user behavior, text, image and location, to find similar items, as shown in Figure \ref{fig:overview}. The individual features are created in the following way:

(i) The matrix factorization features are trained with the industry standard Alternating Least Squares (ALS) \cite{spark:als} with a rank of 100 on user behaviors from a 20-day look-back. The model use ad views and sale conversion signals such as contacting seller as input, and the more indicative conversion signals are given more weight. It provides an item presentation based on user behaviors.

(ii) The textual features are from the unstructured free text in the ad title and description. We first pre-train an item classifier based on the ad category chosen by the sellers, and then extract the top layer of this classifier as our textual embedding. The classifier model is a variant of the Convolutional Neural Network (CNN) architecture \cite{collobert2011natural}, for which we use a word2vec \cite{mikolov2013word2vec} model trained on marketplace texts.

(iii) The image features are extracted from the penultimate layer in a pre-trained \textit{Inception-v3} \cite{szegedy2016rethinking} CNN. We then project the image features into the same space as the textual embeddings by training an image-based classifier to predict the mean word2vec embeddings of the item title. The classifier is a dense network with seven layers, trained to minimize the mean squared error between the predicted embeddings and the true embeddings. We use this model to generate a 100-dimensional image representation for the hybrid model.

(iv) We use a trained location representation instead of simple geographical distance, as the simple distance can be misleading. There are hidden factors such as population density and ease of transport that can affect the impact of distance. We train the location embedding based on the historical user behavior, since the items an user showed interest in implicitly tell us a lot about these factors. Similar to training the user-item matrix factorization features, we factorize a user-postcode matrix and use the resulting postcode embeddings as the location representation.

On top of the above four modules, we build a hybrid model that concatenates all the item features together and then transforms the high dimensional features into a 100-dimensional item representation space to calculate the item similarity score. The different item representations are combined into a global item representation to predict which pairs of items a user will convert on the same day. Based on the assumption that a user is likely to be looking for related items on the same day, we use the item pairs they perform conversion actions on the same day as the positive labels for recommendation. We use negative sampling to produce item pairs that are unlikely to attract the same user's attention.

\section{Five successful tactics}
\label{sec:tactic}

We experimented with different features, training strategies and hyper parameter choices to optimize the performance of the hybrid model on \textit{FINN.no}. Our experiment system consists of both offline and online evaluation. We use the offline metric Hit Rate@n (HR@n) \cite{Deshpande04} as a proxy for online conversion rates to evaluate whether a recommender model is feasible for online testing. The models that perform better or in the same range as the baseline in offline tests are deployed to online A/B tests. We measure the click-through rate (CTR) based on those in-view events to evaluate model performance. In this section, we will share the five tactics that demonstrated most significant improvements in the A/B tests. 

\subsection{Rich user behavior signals}
\label{sec:behavior}

A large portion of the ads on \textit{FINN.no} give items away for free, and lots of transactions are completed with cash payment in person. As a result, we cannot track the completion of all transactions accurately. View of ads is often used as a proxy of transactions. For example, in A/B testing, we use click-through-rate on the recommendations to measure the success of the recommenders, which is measuring the conversion rate to view of ads. However, we find it more effective to leverage all signals such as sending messages and obtaining the phone number of the seller that hint towards transactions. When training the Collaborative Filtering (CF) models, we use all the implicit signals in Table \ref{tab:cf} after filtering out users and items with only one single event. 

Including rich user behavior signals not only increases the training data size, but also improves the signal quality by allocating more weights to the signals closer to transactions. We experimented both with using matrix factorization \cite{sarwar2001cf} as a standalone personalized recommender and using its item embeddings as one module of the hybrid model. The models using rich signals always perform better than the ones using view ad signal only. 

Note that when training the hybrid model, training data size is no longer an issue as we are not dealing with the user-item matrix in this final step. Label accuracy becomes more important than pure volume, so we use only two signals - sending message and viewing phone number - to train the hybrid model. They are the most accurate signals, for they are the closest actions to transactions on \textit{FINN.no}. 

\begin{table}
  \caption{Signals used in collaborative filtering.}
  \label{tab:cf}
  \begin{tabular}{p{0.25\linewidth}|p{0.7\linewidth}}
    \toprule
    \textbf{Feature} & \textbf{Description}\\
    \hline
    \texttt{View ad} & View the ad detail page \\
    \hline
    \texttt{Show interest} & Click on the "read more" link at the end of the ad (only available for some categories) \\
    \hline
    \texttt{Follow seller} & Subscribe to the email alert of new ads from the same seller \\ 
    \hline
    \texttt{Favorite ad} & Add the ad into a favorite item list \\
    \hline
    \texttt{Send message} & Contact the seller using the in-product messaging feature \\
    \hline
    \texttt{Show phone} & Obtain seller's mobile number \\ 
    \hline
    \texttt{Contact seller} & Contact the seller using phone call or SMS (only available in mobile applications) \\ 
    \bottomrule
  \end{tabular}
\end{table}

\subsection{User behavior trained embeddings}
\label{sec:embedding}

To create a suitable representation of the textual features, we use the user-input item category and subcategory as the training data. We first train a classifier to predict the category and subcategory from item title and description, and then extract the internal representations of this classifier as our textual embedding. The classifier model we use is a variant of the Convolutional Neural Network (CNN) architecture of Collobert et al. \cite{collobert2011natural}, where we use a word2vec \cite{mikolov2013word2vec} model trained on marketplace texts for word embeddings. 

Moreover, using behavior as the training data not only captures more detailed item type than the category system, but also catches the common co-click patterns between different items. We observed some examples such as horseback riding and dog, baby clothes and baby toy in the A/B tests, and later confirmed users' positive attitude towards them in user studies. The item embedding from collaborative filtering carries these information and performs well with items with clicks. 

In addition to the item and textual embeddings, we use user behavior signals to train the location embedding too. Representing the location of an ad presents special challenges. In densely populated regions such as cities, a buyer might not be willing to travel that far for an item they are interested in, whereas in sparsely populated regions such as the countryside, a buyer might be willing to travel much farther. Furthermore, locations that seem geographically close to each other might be separated by mountain ranges or other obstacles, making direct transport difficult. Thus pure geographical distance cannot accurately reflect the location impact on conversion potential. We need a representation that takes into account hidden factors such as population density and ease of transport. To achieve this we train a location embedding based on the behavior of our users, since the ads an user has shown interest in implicitly tell us a lot about these factors. We use standard ALS as in Section \ref{sec:behavior}, but instead of a user-item matrix, we factorize a user-postcode matrix, and use the resulting postcode embedding as location representations.

In the A/B test, we compare models using all three types embeddings to models using only content-based textual embedding.

\subsection{Transfer learning on images}
\label{sec:image}

For textual data, the features from the classifier pre-trained with the category and subcategory of items serves well as the textual embedding. However, using category and subcategory as labels does not work as well for image features, because image data has much higher dimensions. As a result, we generate labels with the textual embedding instead. We use the same word2vec \cite{mikolov2013word2vec} model trained on marketplace texts for word embeddings. We then project the image features into the same space as our text embeddings by training a classifier to predict the mean word2vec embeddings of the first five words in the title of an item. The image classifier uses the true title embedding as label to train to minimize the mean squared error between the predicted embeddings and the true embeddings.
%
%
%
%

However, when we tried to train a deep CNN architecture from scratch in this way, we struggled to get good results due to the sparse and imbalanced data distribution. In order to solve this, we use transfer learning to ensure that small classes with only few samples are also well represented.

Instead of training from scratch, we first extract the image features from the penultimate layer in an \textit{Inception-v3} \cite{szegedy2016rethinking} CNN architecture pre-trained on the \textit{ImageNet} \cite{ILSVRC15} classification task, resulting in a 2048-dimensional feature vector. It serves as a reasonable baseline, and we only need to a small dense network with seven layers with the textual embeddings to generate 100-dimensional image embeddings for the hybrid model. We observed moderate CTR improvement by adding the image embeddings to the hybrid model.

\subsection{Staged training strategy}
\label{sec:staged}

End-to-end training has become common recently due to its advantages of less feature engineering and concise system design. However, we chose a staged training strategy in our particular application for three reasons: limited volume of training data, performance stability concern and computing cost.

As mentioned in Section \ref{sec:behavior}, we only use high quality conversion signals to train the hybrid model. Those signals are relatively rare (one magnitude lower than ad views) and they cannot satisfy the massive training data needs of the end-to-end training strategy. Therefore we leverage more noisy signals such as ad views and item category to pre-train the modules in the first stage, and only use the conversion signals to train the "shallow" hybrid part in the later stage. 

Deep models trained end-to-end are difficult to interpret. For recommenders, we do not have a golden dataset that can accurately measure their performances before putting them into online experiments. We have observed several cases that improved the Hit Rate@n offline but underperformed the baseline later in the online experiments. We refresh the recommenders in production daily with the latest user behaviors. Even though the model architecture is stable, there can be significant weight changes. The staged training strategy allows us to have a certain level of performance insurance by analyzing the outcomes of each module.

Computing cost is also a practical reason for choosing the staged training strategy. We share the modules among several search and recommendation features. By training in a staged fashion, we can reuse the modules from the first stage. This shortens the computing time of the daily model refresh as well. The fresher models enable new items to appear in the recommendations sooner, which is particularly important for marketplaces as explained in Section \ref{sec:intro}.

\subsection{Attention mechanism}
\label{sec:attention}

The key of solving the cold start problem is to pick up the content features for fresh items while leveraging user behavior features whenever they are available. This challenge is handled in the hybrid model of the second stage. We introduced the attention mechanism here to achieve the shift of focus.

\begin{figure}
\centering
\includegraphics[width=\linewidth]{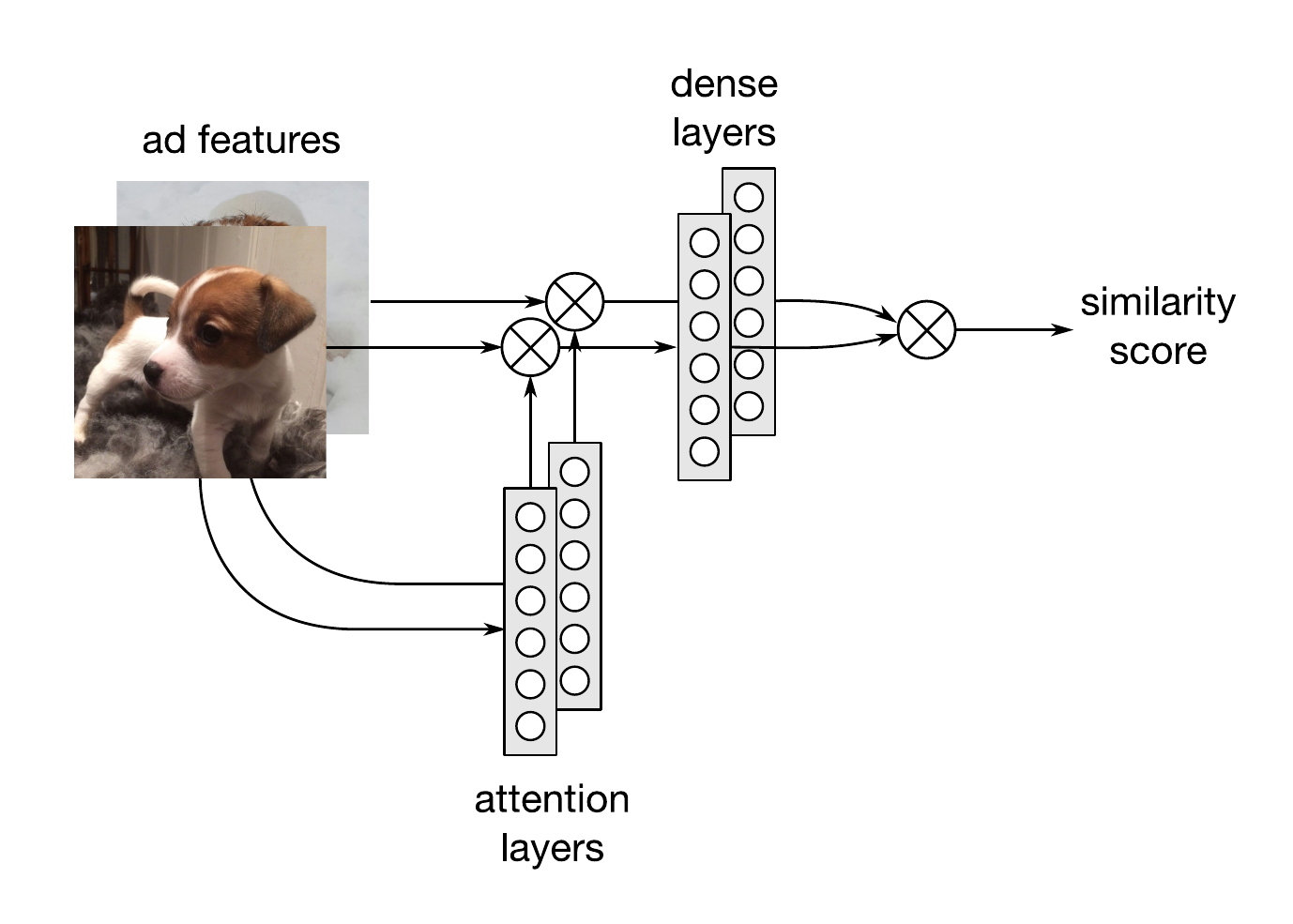}
\caption{The hybrid model.}
\label{fig:ensemble}
\end{figure}

The hybrid model is a Siamese network, taking item representations computed in the first stage as input as shown in Figure \ref{fig:ensemble}. All the item features are concatenated and passed through an attention layer. Then, a towering feed forward network is used to transform the high dimensional features into a 100-dimensional item representation (which is a manageable dimension for our serving infrastructure). Two items are compared using cosine similarity. 

\begin{figure}
\centering
\includegraphics[width=0.8\linewidth]{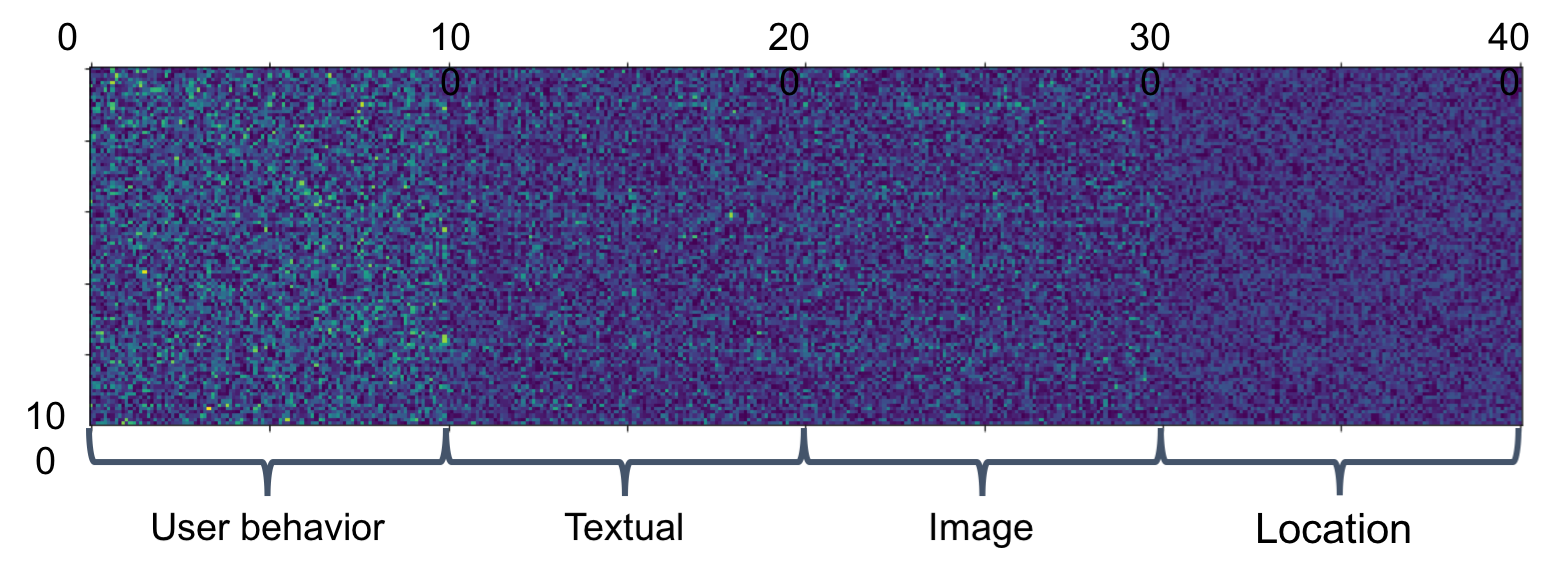}
\caption{The visualization of the weights in the hybrid model.}
\label{fig:ensemble-weight}
\end{figure}

We visualized the weight of the final dense layer of the towering feed forward network in the hybrid model in Figure \ref{fig:ensemble-weight}. We can see that user behavior features still play the most important role, but content based features clearly help as well. Thanks to the attention layer, when the item representations from collaborative filtering are missing, the hybrid model can focus on the content based features to still output a reliable item similarity score, and vice versa. We compared this model to a simple linear hybrid model in the A/B test, and the CTR improvement is quite significant.

\section{Results}
\label{sec:results}

\textit{FINN.no} has over one million active ads in 10+ categories and 200+ subcategories for sell. It offers about 10 million clicks per day that are used as our training data. We tested the above-mentioned five tactics incrementally in production with a recommendation widget at the bottom of an ad detail page. We fix the user interface to displaying six recommended items. Clicking on any of the items count as a click of the recommendation. By default, the A/B tests last for one week to avoid seasonal impact of weekends and accumulate around one million page views in which users did view the recommendation widget. The final model after parameter tuning achieved 22.3\% CTR on \textit{FINN.no}, which has improved about 50\% over the off-the-shelf matrix factorization.

The results are summarized in Table \ref{tab:results}. As we did not experiment all models in the same A/B test, the absolute click through rates (CTR) varies a lot due to the seasonal effect, new feature launches, etc. Hence, we report the more stable CTR improvement $\Delta CTR = (CTR_B - CTR_A) / CTR_A$ when comparing results. Based on the results, transfer learning on images and rich user behavior signal were less impactful than the other three. For the image features, we are investigating the accuracy and sparsity of the labels to see if we can further improve the performance. For user behavior, it can be due to the limited volume of rich behavior signals, which is orders of magnitude lower than the basic ad view volume.

\begin{figure*}
\centering
\begin{subfigure}{\textwidth}
\centering
	\includegraphics[width=0.75\linewidth]{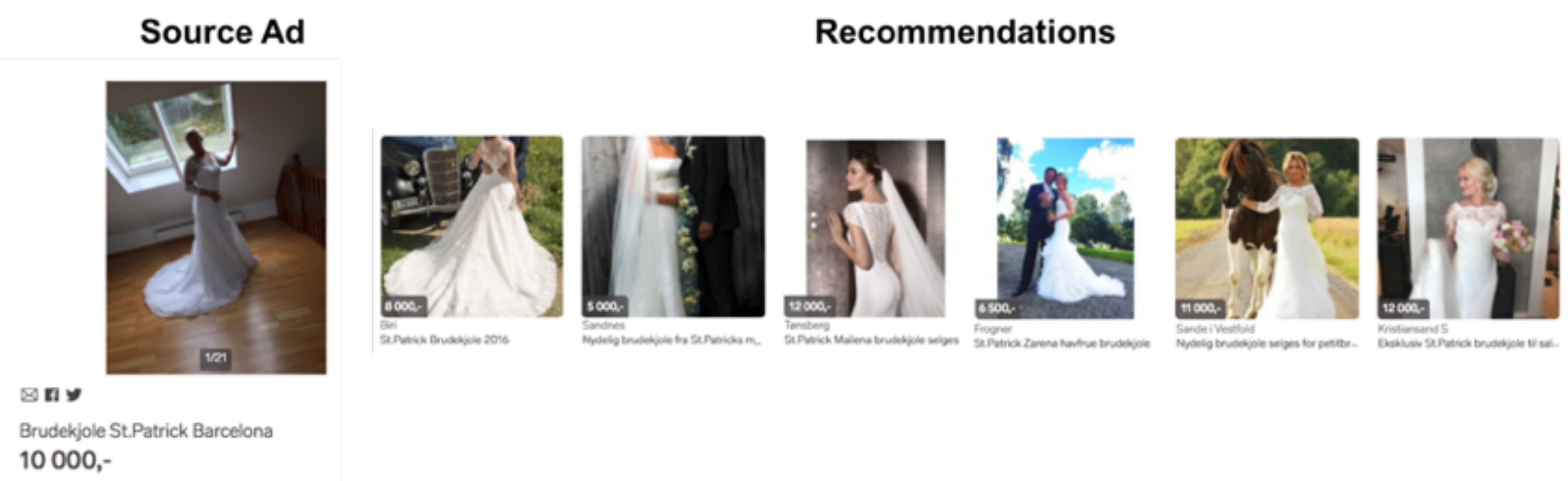}
    \caption{Clothes, cosmetics and accessories}
\end{subfigure}
\begin{subfigure}{\textwidth}
\centering
	\includegraphics[width=0.75\linewidth]{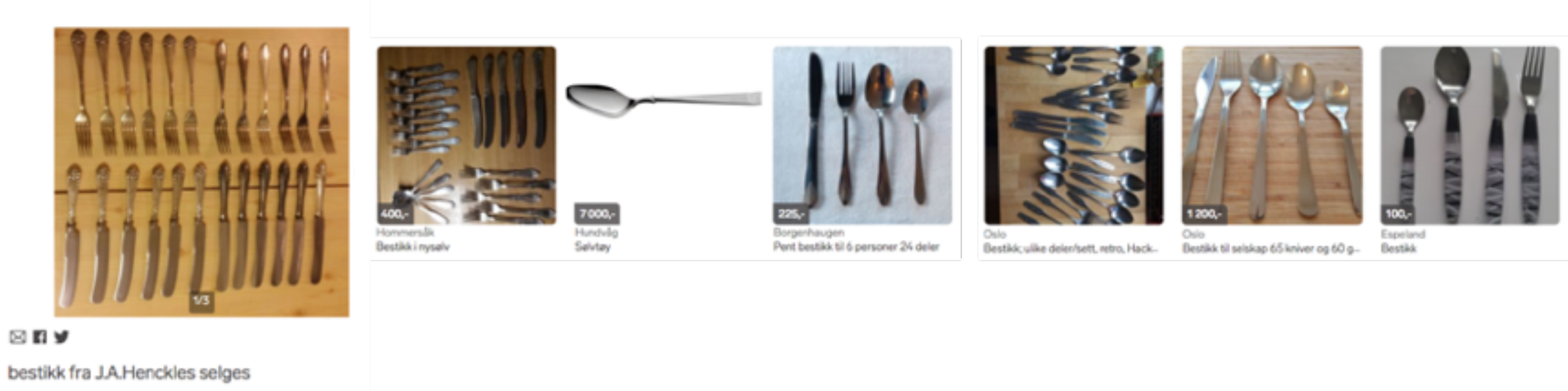}
    \caption{Furniture and interior}
\end{subfigure}
\begin{subfigure}{\textwidth}
\centering
	\includegraphics[width=0.75\linewidth]{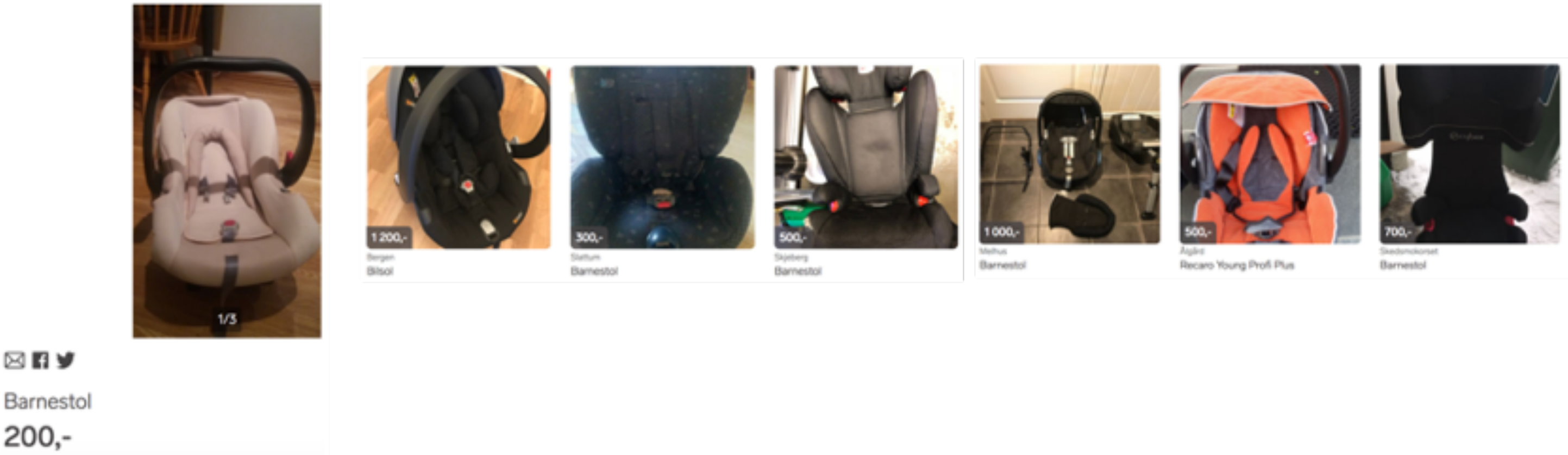}
    \caption{Parents and children}
\end{subfigure}
\begin{subfigure}{\textwidth}
\centering
	\includegraphics[width=0.75\linewidth]{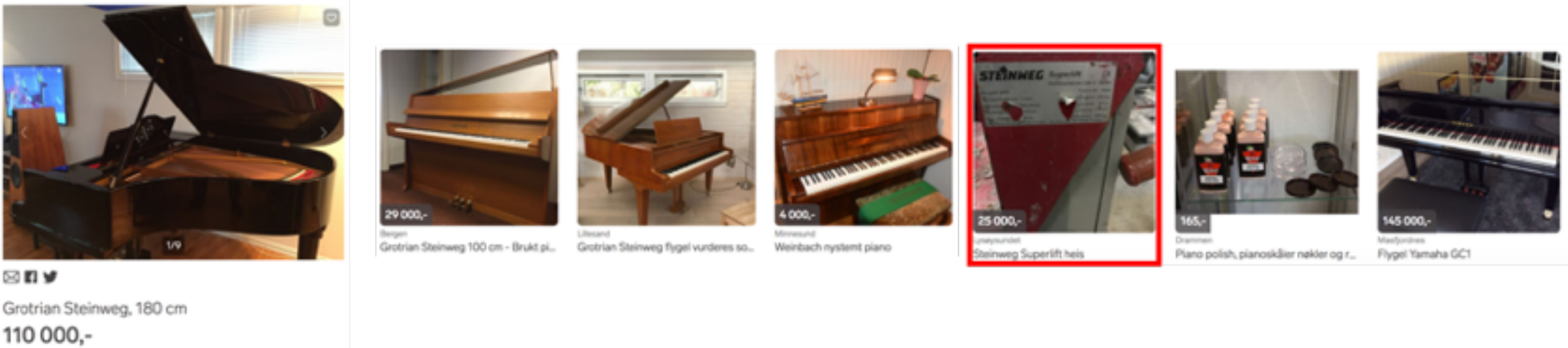}
    \caption{Leisure, hobby and entertainment}
    \label{fig:results:leisure}
\end{subfigure}
\begin{subfigure}{\textwidth}
\centering
	\includegraphics[width=0.75\linewidth]{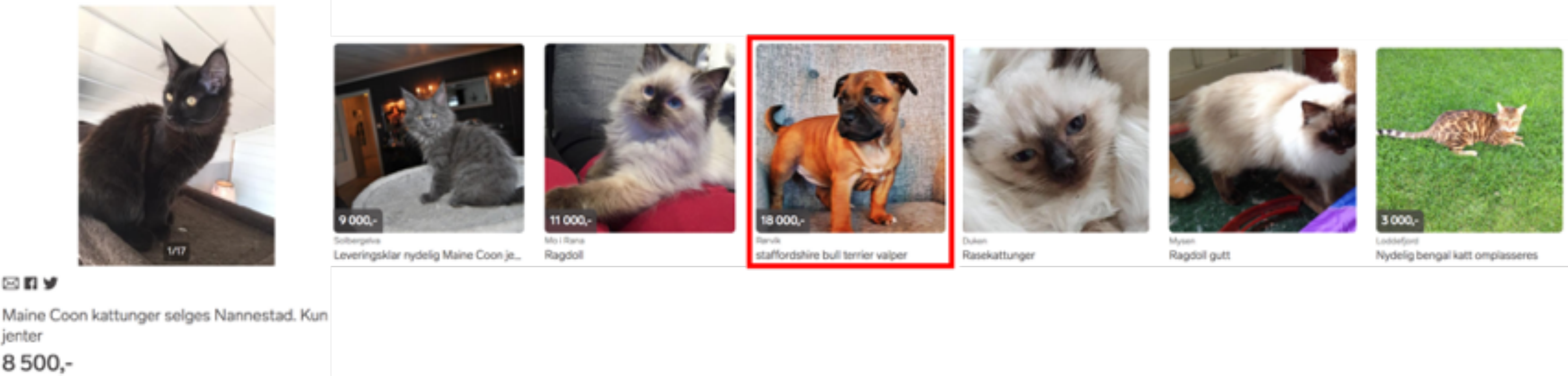}
    \caption{Animal and equipment}
\end{subfigure}
\caption{Sample recommendations for ads from different categories.}
\label{fig:results}
\end{figure*}

Sample results for different categories are shown in Figure \ref{fig:results}. Poor recommendations are marked in red. They often occur on cold-start items that accidentally match with high scores in some textual or image features. For example, for the Steinweg piano in the Leisure, hobby and entertainment category in Figure \ref{fig:results:leisure}, a Steinweg-branded elevator was wrongly recommended. We consider adding more explicit features such as category and brand to the content part of the model to solve this better.

Moreover, the pattern of user interest change is far more complex than only focusing on the fresh items. By examining the categories users visit, we notice that they can have a combination of a \textit{short intent}, i.e. similar to the intent of their last visit, and a \textit{long intent}, i.e. recurring across several visits with a longer "memory". Users' intent might drift during the same visit. This can be triggered by switching intent or updates from a certain item. There are still lots of information for our recommenders to discover from the user behavior data, and this will be our main focus in the next step.

\begin{table}
  \caption{Experiment Results. Please refer to subsections in Section \ref{sec:tactic} for the details of the baseline and experimental models.}
  \label{tab:results}
  \centering
  \begin{tabular}{c|c}
    \toprule
    \textbf{Tactic} & \textbf{$\Delta$CTR} \\
    \midrule
    rich user behavior signal & 7.7\% \\
    \hline
    user behavior trained embeddings & 20.5\% \\
    \hline
    transfer learning on images & 2.1\% \\
    \hline
    attention mechanism & 11.5\% \\
    \bottomrule
  \end{tabular}
\end{table}

\section{Conclusions}
\label{sec:conclusion}
In this paper, we present five tactics that work well with a hybrid recommender model that combines user behavior with textual, image and location features to provide ad recommendations for a marketplace. The model uses rich user behavior signals to extract item similarity features from collaborative filtering and combines them with content-based embeddings to solve the cold-start problem. It also leverages user behaviors as labels and employs transfer learning to capture detailed similarity that is not easily described by the user-selected categories. The recommender overcomes the challenges of heterogeneous data with a staged training strategy and demonstrates improved click-through rate in experiments with real users.

In the future, we plan to introduce more effective models as modules into the hybrid model. For example, sequence-based models such as ATRank \citep{atrank:2018} work particularly well as a standalone recommender, because they can utilize the sequential information in user behaviors to handle the multi-intent and intent shift. We want to combine them as a module into the hybrid model as well. Moreover, though the staged training strategy works well at the current stage, we want to investigate if we can go more end-to-end to streamline the architecture of the recommender.

\begin{acks}
The authors would like to thank Nicola Barbieri and Thorkild Stray for their valuable comments and helpful suggestions.
\end{acks}